\documentclass[twocolumn,secnumarabic,amssymb, nobibnotes,aps,pra]{revtex4-1}
\usepackage[english]{babel}

\usepackage[normalem]{ulem}
\usepackage[dvipsnames]{xcolor}
\usepackage{float}
\usepackage{textcomp}

\usepackage{amsfonts}
\usepackage{amsmath}
\usepackage{graphicx}
\usepackage{color}
\usepackage{hyperref}

\begin{document}



\title{Mode-locking induced by coherent driving in fiber lasers}
\author{Carlos Mas Arabí$^1$, Nicolas Englebert$^1$, Pedro Parra-Rivas$^{1,2}$, Simon-Pierre Gorza$^1$, and François Leo$^1$}
\affiliation{$^1$OPERA-photonics$,$ Université Libre de Bruxelles$,$ 50 Avenue F. D. Roosevelt$,$ CP 194/5 B-1050 Bruxelles$,$ Belgium}
\affiliation{$^2$Dipartimento  di  Ingegneria  dell\textquotesingle Informazione, Elettronica  e  Telecomunicazioni,
Sapienza  Universit\`a  di  Roma, via  Eudossiana  18, 00184  Rome, Italy}

\begin{abstract}
The generation of stable short optical pulses in mode-locked lasers is of tremendous importance for many applications. 
Mode-locking is a broad concept that encompasses different processes enabling short pulse formation. 
It typically requires an intracavity mechanism that discriminates between single and collective mode lasing, which can be complex and sometimes adds noise. Moreover, known mode-locking schemes do not guarantee phase stability of the carrier wave.
Here we theoretically propose that injecting a detuned signal seamlessly leads to mode-locking in fiber lasers. 
We show that phase-locked pulses, akin to cavity solitons, exist in a wide range of parameters. In that regime the laser behaves as a passive resonator due to the non-instantaneous gain saturation.
\end{abstract}

\maketitle

Optical Dissipative Solitons (DSs) are pulses propagating without distortion in an optical cavity~\cite{akhmediev_dissipative_2005}. DSs belong to the wider class of dissipative localized structures which emerge in different fields such as hydrodynamics~\cite{wu_observation_1984} or plasma physics~\cite{Kim_cavitons_74}.
DSs can take many shapes, depending on the parameters of the system. We focus on the dissipative counterparts of the well-known sech-shaped nonlinear Schr\"odinger soliton, which attracts a lot of attention in both mode-locked lasers~\cite{grelu_dissipative_2012} and passive nonlinear resonators~\cite{Wabnitz:93,leo_temporal_2010,herr_temporal_2014}. In the latter system, they are called cavity solitons (CSs).
The main difference between lasers and passive resonators lies in how the energy is provided to the system. In lasers, the gain is incoherent while in passive resonators energy comes from an external coherent driving.
The advantage of coherent driving is that it adds an important control parameter through the cavity detuning, leading to bistability in Kerr resonators~\cite{lugiato_spatial_1987}, which in turn allows for the formation of coherent solitons on a stable background~\cite{scroggie_pattern_1994}.
In lasers, solitons are not phase-locked and they are only stable on the condition that the trivial off solution is stable in between pulses~\cite{haus_theory_1975}, which requires active or passive mode-locking mechanisms such as intracavity modulation~\cite{hargrove_locking_1964}, saturable absorbers~\cite{haus_theory_1975}, and Kerr lensing~\cite{spence_60-fsec_1991} among others ~\cite{ippen_additive_1989,wright_spatiotemporal_2017,liu_megawatt_2017,bao_laser_2019}. 

Here we theoretically show that an injected continuous wave (cw) signal seamlessly leads to mode-locking in fiber lasers, through the formation of ultrastable solitons, without the need for any additional intracavity mechanism.  In connection to our recent results on soliton formation in active resonators pumped below the lasing threshold~\cite{englebert_temporal_2021}, we call them active cavity solitons (ACSs).
ACSs exist in the regime where the \emph{saturated} incoherent gain is lower than the intracavity loss. In that configuration, the laser cavity can be treated as a low-loss passive resonator. 
ACSs are hence intrinsically linked to CSs. They are phase-locked to a driving laser which forms a homogeneous background around the sech-shaped soliton.
Laser solitons can also be phase-locked to a continuous wave driving signal in a configuration often called injection locking~\cite{margalit_injection_1996,rebrova_stabilization_2010,komarov_nature_2014}.
We demonstrate that the role of injection goes beyond adding coherence. It may induce mode-locking in the absence of standard schemes and the detuning can be harnessed to tune the pulse properties.

\begin{figure}[H]
    \centering
    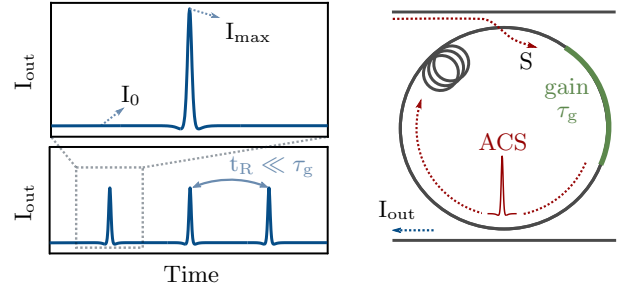

    \caption{Schematic representation of a coherently-driven active Kerr fiber resonator. The cavity round-trip time ($t_R$) is several orders of magnitude shorter than the gain lifetime ($\tau_g$).}
    \label{fig:dispositif}
\end{figure}

The physical system we consider is depicted in Fig.~\ref{fig:dispositif}. It consists of a driven fiber resonator incorporating a short erbium-doped fiber amplifier. 
Under some conditions, in particular when the gain dynamics is much slower than the round-trip time, the dimensionless slowly varying electric field envelope $E$ and gain $g$ can be modeled by the following normalized mean-field model: 
\begin{align}
    &\dfrac{\partial E(T,\tau)}{\partial T} = \bigg[ -1+
    g(T) +i(|E(T,\tau)|^2-\Delta)  \label{eq:meanfield_normalized} \\\nonumber
    &-i\eta\dfrac{\partial^2}{\partial\tau^2}
    \bigg]E(T,\tau)+S,\\
    &\dfrac{d g(T)}{d T} = \mu [-(1+\xi \langle|E(T,\tau)|^2\rangle)g(T)+\mathcal{G}_0],
    \label{eq:gain_normalized}
\end{align}

\begin{figure}
    \centering
    \includegraphics{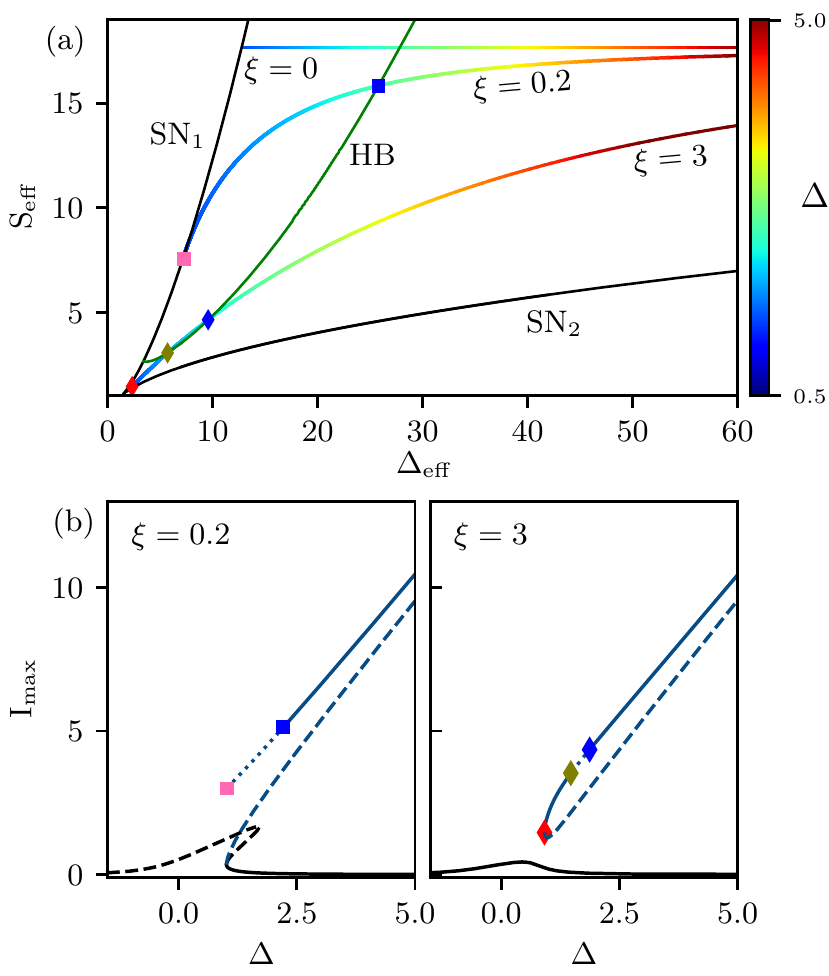} 
    \caption{(a) Phase-space ($S_{\text{eff}}$, $\Delta_{\text{eff}}$) of the Lugiato-Lefever equation. Soliton existence is delimited by the saddle-node bifurcations SN$_1$ and SN$_2$ (black lines). The Hopf Bifurcation (HB) leading to soliton spatiotemporal oscillations is represented in green. 
    The rainbow line shows the trajectory of our system for different saturation powers as we change the detuning $\Delta$. (b) Soliton peak power (blue line) and power of the cw solutions (black) of the active cavity as a function of $\Delta$. Solid line: stable solutions. Dashed line: saddle solutions. Dotted line: Hopf unstable solutions. The markers indicate the position of the corresponding crossings of bifurcation lines in panel (a). Parameters: $t_c=
    140,000$, $\mathcal{G}_0=0.92$ and $S=0.4$.}
    \label{fig:rainbow}

\end{figure}

where $T$ is the slow time scaled with respect to the round-trip time $t_R$, $\tau$ is  time in a reference frame traveling at the carrier frequency group velocity, $\Delta$ is the normalized phase detuning, $S$ is the normalized driving, $\mathcal{G}_0$ is the ratio between the small-signal gain and the intrinsic cavity loss, 
$\xi$ is related to the saturation power, and $\mu=\alpha_it_R/\tau_g$ where $\tau_g$=10 ms is the erbium relaxation time.
The normalized parameters are linked to physical quantities through the relations: $T\rightarrow \alpha_i T/t_R$  where $\alpha_i$ is equal to the total intrinsic cavity losses; $\Delta=\delta/\alpha_i$, where $\delta$ is the cavity phase detuning in physical units; $\tau\rightarrow \tau \sqrt{2\alpha_i/(|\beta_2|L)} $ where $\beta_2$ is the group velocity dispersion; $\eta=\mathrm{sign}(\beta_2)$ and $L$ is the cavity length; $E\rightarrow E\sqrt{\gamma L/\alpha_i}$, where $\gamma$ is the Kerr nonlinear coefficient, $S^2=\frac{P_{in}\gamma L \theta}{\alpha_i^3}$ where $\theta$ is the power transmission coefficient of the coupler; $\xi= \alpha_i/(\gamma L P_{sat})$, where $P_{sat}$ is the saturation power. 
The average power $\langle|E|^2\rangle$ is evaluated over one roundtrip~\cite{haboucha08,niang_influence_2015}  : $\langle|E|^2\rangle=t_c^{-1}\int_{-t_c/2}^{t_c/2}|E|^2d\tau$~\cite{haboucha08,niang_influence_2015} where $t_c=t_R\sqrt{2\alpha_i/(|\beta_2|L)}$ is the normalized round-trip time. 
In what follows, we focus on the anomalous regime ($\eta=-1$). 


The stationary solutions of Eqs~\eqref{eq:meanfield_normalized}-~\eqref{eq:gain_normalized} satisfy the equation 
\begin{align}
0 = \bigg[ -1+ \frac{\mathcal{G}_0}{(1+\xi \langle|E(\tau)|^2\rangle)} +i(|E(\tau)|^2&-\Delta) + \nonumber \\
    +i\dfrac{\partial^2}{\partial\tau^2}
    &\bigg]E(\tau)+S. \label{eq:LLE}
\end{align}
We readily note that this equation resembles the stationary Lugiato-Lefever equation (LLE)~\cite{lugiato_spatial_1987}. The only difference comes from the additional saturated gain term. 
In this work, we focus on the region where the saturated gain is lower than the intracavity loss such that the cavity behaves as a passive resonator with high effective finesse. 
For solitons hosted in long passive resonators, most of the optical energy stored in the resonator comes from the low-power cw background.
We start by making the approximation $\langle|E(\tau)|^2\rangle\approx I_0$, where $I_0$ is the power of the cw background (see Fig. \ref{fig:dispositif}), to identify the regions of existence of cavity solitons in our system. By introducing the effective loss $\alpha_{\text{eff}}=1-\mathcal{G}_0/(1+\xi I_0)$ in Eq.~\eqref{eq:LLE}, one recovers the LLE describing a passive resonator with total round-trip loss $\alpha_{\text{eff}}$~\cite{lugiato_spatial_1987,haelterman_low_1992}. The oft-used dimensionless driving ($S_{\text{eff}}$) and detuning ($\Delta_{\text{eff}}$) parameters of the LLE~\cite{Coen_13_universal} can be retrieved through the relations $S_{\text{eff}} = S/\alpha_{\text{eff}}^{3/2}$ and $\Delta_{\text{eff}}=\Delta/\alpha_{\text{eff}}$, when $\alpha_{\text{eff}}>0$.
For every set $(S,\Delta,I_0$), where $I_0$ is a solution of 
\begin{equation}
    S^2=I_0\left[\left(1-\frac{\mathcal{G}_0}{1+\xi I_0}\right)^2+(\Delta-I_0)^2 \right],
    \label{eq:cw_solution}
\end{equation}
we can calculate the effective LLE parameters ($S_{\text{eff}}$,$\Delta_{\text{eff}}$) and predict the existence of solitons and their stability in our active system.
The region of existence of solitons in the LLE is well known~\cite{parra-rivas_dynamics_2014,Coen_13_universal}. In the $(S_{\text{eff}},\Delta_{\text{eff}})$ space, they are located in the region bounded by the saddle-node bifurcations SN$_1$ and SN$_2$ [see Fig.~\ref{fig:rainbow} (a)]. SN$_1$ marks the low cw fold of the LLE and SN$_2$ is the soliton saddle-node. The latter is well approximated by the expression SN$_2=2\alpha_{\text{eff}}\sqrt{2\Delta}/\pi$~\cite{nozaki_low-dimensional_1986,Wabnitz:93}.
The Hopf bifurcation (HB) line  (calculated numerically), indicates the region where solitons lose stability and oscillatory behavior, as well as spatiotemporal dynamics, can be found~\cite{Leo:13,Anderson:16,Pedro18PRE}.

As examples of trajectories in the $(S_{\text{eff}},\Delta_{\text{eff}})$ plane, we use parameters which correspond to our recent experimental results~\cite{englebert_temporal_2021}.
In that configuration, the small-signal gain is lower than the intracavity loss (no lasing) and $(S_{\text{eff}},\Delta_{\text{eff}})$ can be defined for all detunings $\Delta$. We will then generalize the concept by showing that similar solitons emerge above the lasing threshold.
Three different $\Delta$-parametrized paths, corresponding to different saturation powers are shown in Fig.~\ref{fig:rainbow}(a). The fixed parameters are $S=0.4$, $\mathcal{G}_0=0.92$ and $t_c=140,000$.
For large detunings ($\Delta>5$, not shown), the background power $I_0$ is low and all trajectories (increasing $\Delta$) asymptotically approach $S_{\text{eff}}=S/(1-\mathcal{G}_0)$, which corresponds to the normalized driving amplitude of a cavity with non-saturable gain ($\xi=0$).
Solitons are predicted to exist up to $\Delta = 30$ where $S_{\text{eff}}=17.7$.

Below $\Delta=5$, gain saturation impacts the effective loss and the trajectory (with decreasing $\Delta$) bends downward. The bend depends on the saturation power. 
For $\xi=0.2$, the bend is weak. The system crosses the HB, leading to oscillatory dynamics and the branch terminates at SN$_1$. 
For lower saturation powers ($\xi=3$), the downward bend is stronger and the system crosses the bottom saddle-node SN$_2$, here at $S_{\text{eff}}=1.27$, showing that a saddle and stable soliton connect for the second time. The HB line is crossed twice, indicating a smaller region with oscillatory states.

To confirm these predictions, we calculate the bifurcation structure of the full model [Eqs.\eqref{eq:meanfield_normalized}~and~\eqref{eq:gain_normalized}].
We use a standard numerical continuation algorithm (the open distribution software AUTO-07p~\cite{Doedel07auto-07p:continuation2}). The solutions are calculated in the domain $\tau=[0,t_c/2]$ using Neumann boundary conditions~\cite{champneys_numerical_2007}. $\langle|E|^2\rangle$ is obtained through an additional integral constraint and is treated as a free parameter. The stability is calculated by computing the  eigenvalues of the Jacobian  matrix associated with Eqs.\eqref{eq:meanfield_normalized}~and~\eqref{eq:gain_normalized}.
The cw and soliton solutions corresponding to the trajectories of Fig. ~\ref{fig:rainbow}(a) are shown as a function of the detuning in Fig.~\ref{fig:rainbow}(b).

For $\xi=0.2$, the cw resonance is bistable, albeit on a small detuning interval and a Turing instability is present at $\Delta=-1.18$. The soliton branch emerges from the lower cw fold and is unstable up until $\Delta=30$ (not shown), where the stable soliton branch is created in a saddle-node bifurcation. For $\xi=3.0$, the cw resonance is single-valued for all detunings and does not undergo modulation instability (MI) [see Fig.~\ref{fig:rainbow}(b)]. Soliton states form two branches, one stable and the other unstable, connected on both ends by a saddle-node bifurcation [at $\Delta$ = 0.9 and $\Delta$ = 30, corresponding to the two crossings of SN$_2$]. This structure is commonly called an isola~\cite{beck_snakes_2009}.
The stable (top) branch undergoes two HBs at low detunings. 
The bifurcation structures of Fig~\ref{fig:rainbow}(b) are in excellent agreement with the predictions inferred from the effective LLE parameters. 

\begin{figure}
    \centering
    \includegraphics{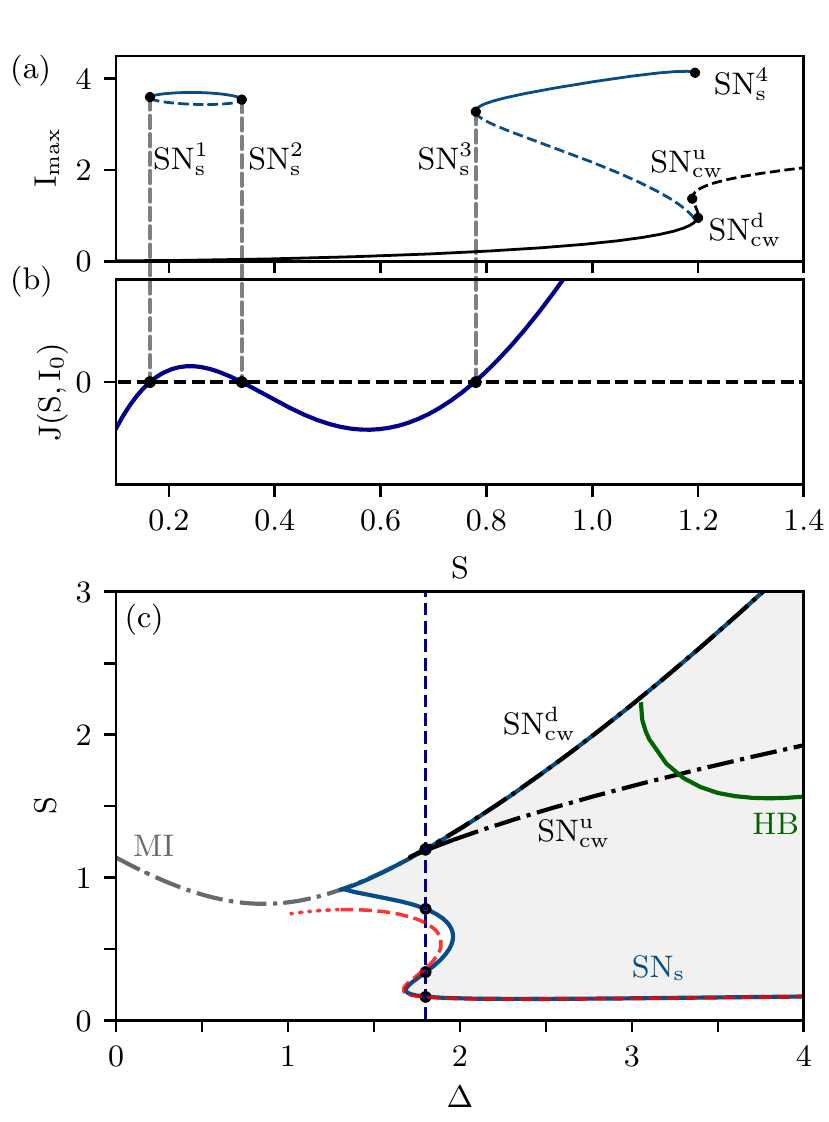}
    \caption{(a) Soliton peak power (blue), and power of the homogeneous states (black) as a function of the driving amplitude $S$ for $\Delta=1.8$ (b) $J(S,I_0)$ as a function of the driving $S$ for $\Delta=1.8$. (c) Phase-space $(\Delta,S)$ of the system showing the soliton (blue) and homogeneous (black) saddle nodes. HB: Hopf Bifurcation (green line). The dashed red line corresponds to the approximation $J(S,(S/\Delta)^2)=0$ of SN$_s$. 
    Parameters: $\xi=7.75$, $\mathcal{G}_0=0.92$ and $t_c=140,000$}
    \label{fig:J}
\end{figure}

We next calculate the bifurcation structure for a lower saturation power ($\xi=7.75$).
We focus on the low-detuning region, where the existence of ACSs is predicted by the condition $S_{\text{eff}}>\text{SN}_2= 2\sqrt{2\Delta_{\text{eff}}}/\pi$, which can be written $J(I_0,\Delta)>0$, where $J(I_0,\Delta)\equiv S-2\alpha_{\text{eff}}\sqrt{2\Delta}/\pi$. 
Interestingly, at this saturation level, there is a small range of detuning where the function $J$ possesses three zeros. This situation  is shown in Fig.~\ref{fig:J}(a), where we plot the soliton branches and $J(I_0, S)$ as a function of the driving amplitude $S$ for $\Delta = 1.8$. At low driving powers, an isola of ACSs is found, corresponding to the first two zeros of $J(I_0, S)$.
The top branch of the isola is stable while the bottom one is unstable (saddle). Increasing $S$, we find a large parameter region without any soliton branch. A third saddle node (SN$_s^3$) is present around $S=0.8$ and a stable and an unstable branch emerge again in a saddle-node bifurcation. In this case, they do not form a isola, instead the unstable branch connects with SN$^d_{\text{cw}}$. This latter structure is very similar to the one found in passive resonators \cite{scroggie_pattern_1994}. This is because, for high background power, the gain is almost fully saturated and one recovers the bifurcation structure of the intrinsic cavity.

The two-parameter bifurcation diagram in the ($S$, $\Delta$)-space  for $\xi=7.75$ is shown in Fig.~\ref{fig:J}(c). 
We see that the two separate regions of soliton existence connect through a necking bifurcation around $\Delta=2$. Beyond this point, solitons exist for a very large region of parameters.
Importantly, in contrast with passive resonators~\cite{parra-rivas_dynamics_2014}, the minimum driving amplitude necessary for soliton formation ($S>$SN$_s^1$) is much lower than that for Turing patterns (arising above the MI or SN$_\text{cw}^\text{u}$ lines).
In the region where the effective loss is low, the cw background power can be approximated by $I_0=S^2/\Delta^2$, simplifying the evaluation of the position of SN$_s$. The agreement between the actual SN$_s$ and its approximation is shown in Fig.~\ref{fig:J}(c). The threshold for soliton formation is well predicted by the approximated saddle node.

\begin{figure}
    \centering
    \includegraphics{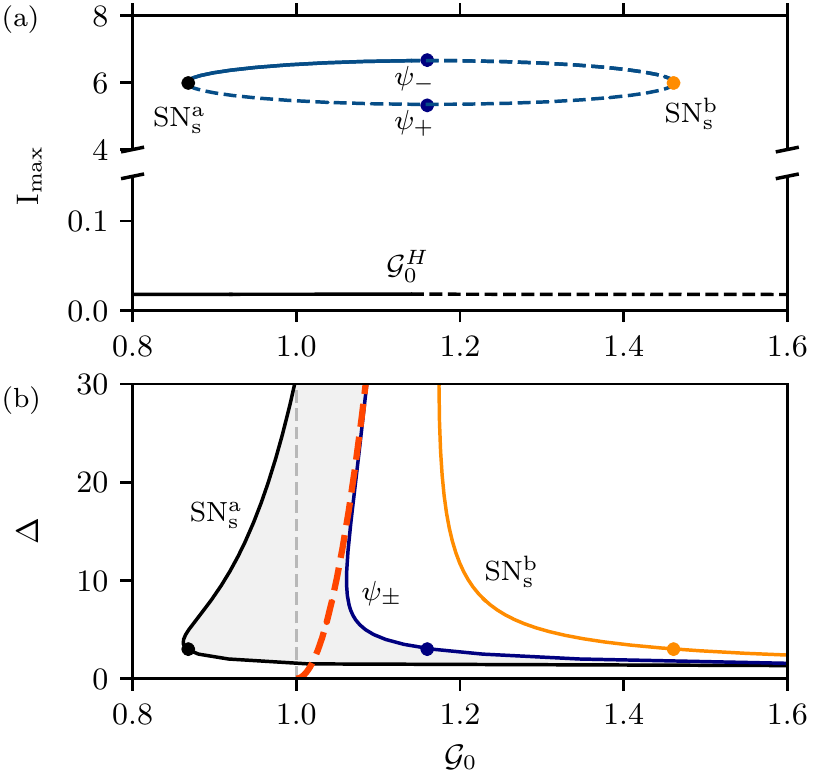}
    \caption{(a) Soliton peak power (blue line), and cw background power (black line) as function of $\mathcal{G}_0$ for $t_c=2000$, $S=0.4$, $\Delta=3.0$, and $\xi=7.75$. Blue dots correspond to $\psi_\pm$. (b) Phase diagram in the $(\mathcal{G}_0,\Delta)$-space showing SN$_s^a$ (black line), SN$_s^b$ (orange line), and $\psi_\pm$ (blue line). The red line corresponds to the nonlinear phase accumulated by the ME soliton. The dashed line at $\mathcal{G}_0=1$ corresponds to the lasing threshold. Stable solitons exist in the grey area.} 
    \label{fig:lasers}
\end{figure}

Finally, we look for connections between ACSs and laser solitons. The latter is the sech-shaped solutions of the master equation (ME) [Eq.~\eqref{eq:LLE} with $S$=0] and they appear above the lasing threshold $\mathcal{G}_0>1$ (see e.g.~\cite{kim_ultralow-noise_2016}).
Because laser solitons are backgroundless, we reduce the normalized round-trip time to $t_c=2000$ in what follows, in order to increase the soliton-to-background energy ratio.
Figure~\ref{fig:lasers}(a) shows a bifurcation diagram as a function of $\mathcal{G}_0$ for $\xi=7.75$, $\Delta=3$ and $S=0.4$. 
The cw solutions are stable until $\mathcal{G}_0^{H}$, where the amplification exactly compensates the intrinsic cavity losses ($\alpha_{\text{eff}}=0$). 
At this point, there is a Hopf bifurcation ~\cite{lugiato_nonlinear_2015}, and modulated (in the slow time) solutions emerge. Further analysis of modulated solutions is beyond the scope of the present paper.
In the region $1<\mathcal{G}_0<\mathcal{G}_0^{H}$, the saturated gain is lower than the intracavity loss, which prevents lasing. This effect is commonly called injection locking  and has been intensely studied in the cw regime~\cite{Tredicce:85,buczek_laser_1973}.
In this region, the effective LLE parameters predict the existence of ACSs. Using these predictions as initial guesses, we compute the soliton solutions as a function of $\mathcal{G}_0$ for a fixed detuning and driving amplitude. The results are shown in Fig.~\ref{fig:lasers}(a).
ACSs form two branches connected at their extremes by SN$_s^a$ and SN$_s^b$. The bottom branch is always unstable, while the upper one is stable up to $\mathcal{G}_0^{H}$.
When the detuning is small, as in Fig~\ref{fig:lasers}(a), the region where stable ACSs can be found is the same as that of stable cw solutions because the gain saturation is mostly set by the cw background. For $\mathcal{G}_0>\mathcal{G}_0^{H}$, the background oscillates as discussed above, and oscillatory ACSs may be found.
The two ACS solutions found at $\mathcal{G}_0=\mathcal{G}_0^{H}$ correspond to the well known analytical solutions of Eq.~\eqref{eq:LLE} for $\alpha_{\text{eff}}=0$~\cite{barashenkov_existence_1996,Matsko:11}. They read:

\begin{equation}
    \psi_{\pm}(\tau)=\psi_0\left[1+\frac{2\text{sinh}^2 \beta}{1\pm \text{cosh} \beta \text{cosh} (B\tau)}\right], \label{eq:Barashenkov} 
\end{equation}

with $\psi_0=(\Delta/(1+2\text{cosh}^2\beta))^{1/2}$, $B=(2\Delta /(1+2\text{cosh}^2\beta))^{1/2}\text{sinh} \beta$, and $\beta$ is a solution of the equation:

\begin{equation}
    S=2\text{cosh}^2\beta\left[\frac{\Delta}{1+2\text{cosh}^2 \beta}\right]^{3/2}.
\end{equation}

These solutions have been discussed in the context of Kerr frequency combs~\cite{Matsko:11}. Here, the solution $\psi_-$ corresponds to the stable soliton with the maximum peak power that can be excited for a given detuning. When the detuning is increased, the soliton peak power increases and the cw background power decreases. In Fig.~\ref{fig:lasers} (b), we show the evolution of the existence region of ACSs, bound by SN$_s^a$ and $\psi_-$, as we change the detuning. 
We compare them to the ME (laser) solitons.
At large detunings, $\psi_-$ and laser solitons asymptotically merge as the cw background of the former tends towards zero.
This connection may highlight the main mechanism behind the concept of soliton injection locking. The sech-shaped solutions of the ME are phase invariant while the $\psi_-$ has a fixed phase relation to the driving laser. Interestingly, there is a very broad region of existence of ACSs beyond these two well-known solutions. When the detuning is large, ACSs are well approximated by the expression~\cite{Coen_13_universal}:
\begin{equation}
    E_{\mathrm{ACS}}\approx\sqrt{2\Delta}\mathrm{sech}\left(\sqrt{\Delta}\tau\right).
\end{equation}
By changing the detuning, one can tune both the duration and the peak power of the solitons. For fixed cavity parameters and driving power, ACSs can be shorter and with higher peak power as compared to the corresponding laser soliton.
Injection locking hence goes beyond fixing the phase of solitons in lasers. It induces the formation of ultra-stable solitons, with an amplitude and duration which can be externally controlled. This mode-locking process has the important advantage that it does not require an additional element such as a saturable absorber~\cite{margalit_injection_1996}. The stability of the system is guaranteed because the gain saturation is larger than the intracavity loss. Conversely, in a laser without injection, the saturated gain is equal to the intracavity loss and an additional mechanism is required to ensure the stability of the pulse in the presence of noise.
Lastly, we recall that coherent driving extends the existence of solitons to regions where laser solitons do not exist, as evidenced in Figure~\ref{fig:lasers}(b) by the large section of stable soliton formation that extends below $\mathcal{G}_0=1$~\cite{englebert_temporal_2021}.

In conclusion, we showed that mode-locking can be obtained through coherent injection in fiber lasers.
Stable solitons (ACSs) exist in a wide region of parameters, which extends both below and above the lasing threshold.
We highlighted their connection to both solitons of passive resonators and lasers, hinting that ACSs may provide the missing link between the two.
These systems are described by equations (the LLE and the ME) which are commonly used in hydrodynamics~\cite{dudley_instabilities_2014} and plasma physics~\cite{nozaki_low-dimensional_1986}, and we expect our findings to extend to these fields.
In future work, we plan to investigate ACS formation in the presence of faster gain dynamics, such as semiconductor optical amplifiers, to bridge the gap with solitons predicted in driven quantum cascade lasers~\cite{columbo_prl_2021}.
Furthermore, the impact of higher-order effects, such as gain dispersion~\cite{haelterman_hopf_1993} the Raman effect~\cite{Wang_2018} or higher-order longitudinal modes~\cite{anderson_coexistence_2017} that may affect soliton formation, especially when the injected signal is strongly detuned from resonance, will be studied. \newline

The authors acknowledge fruitful discussions with Alessia Pasquazi. This work was supported by the European Research Council (ERC) under the European Union’s Horizon 2020 research and innovation program (grant agreement No 757800) and Fonds de la Recherche Scientifique - FNRS under grant No PDR.T.0104.19. P. P. -R acknowledges support from the European Union’s Horizon 2020 research and innovation programme
under the Marie Sklodowska-Curie grant agreement no. 101023717. C.M.A, N.E., P. P.-R and F.L acknowledge the support of the Fonds de la Recherche Scientifique-FNRS

\bibliography{biblioteca2}

\end{document}



\title{Supplementary Material: Dissipative solitons in a driven active Kerr cavity}
\author{Authors}

\maketitle

\section{Analysis of the cw}
In this section, we will go into the details of the cw analysis. We will study which values give a stationary behavior of the gain. Our starting point is:
\begin{equation}
    \dfrac{\partial E}{\partial T} = \bigg[ -1+
    g +i(|E|^2-\Delta) +i\dfrac{\partial^2}{\partial\tau^2}
    \bigg]E+S,
    \label{eq:meanfield}
\end{equation}
where the function $g$ is the gain and depends on the evolution as follows:
\begin{equation}
    \dfrac{\partial g}{\partial T} = \mu [-(1+\xi |E|^2)g+\mathcal{G}_0],
    \label{eq:g}
\end{equation}
where $\mu$ is the ratio between the round-trip time and the time response of the medium. We expand the stationary values of $E=E_0+\epsilon E_1$ and $g=g_0+\epsilon g_1$, where the firsts terms in the expansions are the stationary values. Keeping only the terms at order $\epsilon$, and considering that $\mu\ll 1$, we obtain the equation:
\begin{equation}
    \partial_TE_1=\left[1-g_0+i2|E_0|^2+i\partial_\tau^2\right]E_1+iE_0^2E_1^*.
\end{equation}
Note that, due to the difference between time scales of the gain and the field $E$, the evolution of the perturbations is reduced to one singly equation.  By using the standard ansatz $E_1=a(T)e^{-i\Omega \tau}+b^*(T)e^{-i\Omega \tau}$, we obtain: 
\begin{equation}
    \frac{d}{dT}
    \begin{bmatrix}
    a \\
    b
    \end{bmatrix}
    =
    \begin{bmatrix}
    D(\Omega)+i2|E_0|^2 & iE_0^2 \\
    -iE_0^{*2} & D^*(\Omega)-i2|E_0|^2 
    \end{bmatrix}
    \begin{bmatrix}
    a \\
    b
    \end{bmatrix},
\end{equation}
where $D(\Omega)=\frac{\mathcal{G}_0}{1+\xi|E_0|^2}-1-i(\Omega^2+\Delta)$. The eigenvalues $\lambda_{\pm}$ of the linearized system determine the stability of the solutions. These values can be obtained analytically and are given by the following expression:
\begin{equation}
    \lambda_{\pm}=\mathcal{G}-1\pm \sqrt{I^2-(\Delta-2{\color{black}I}+\Omega^2)^2}, 
\end{equation}

where $\mathcal{G}=\frac{\mathcal{G}_0}{1+\xi|E_0|^2}$, and $I=|E_0|^2$.
\subsection{$\Omega=0$}

For $\Omega=0$, the eigenvalues are given by:

\begin{equation}
    \lambda_{\pm}=\mathcal{G}-1\pm \sqrt{-3I^2+4I\Delta-\Delta^2}, 
\end{equation}
We can identify two bifurcations:
\begin{itemize}
    \item The injection locking point ($\mathcal{G}=1$) , which corresponds to a Hopf bifurcation. Note that, for low values of $I$, Kerr nonlinearity can be neglected and the frequency of the arising oscillations is $\Delta$.
    \item $\lambda_\pm=0$ . This condition leads to two kinds of instability that correspond to two well differentiated physical phenomena: on the one hand, this bifurcation leads to a saddle node of the cw and thus, to the well-known bistability of the cw in the passive limit ($\xi\rightarrow 0$). On the other hand, this bifurcation can also appear in regions where the cw is not multi-valuated. Our numerical show that this bifurcation leads to an oscillatory behavior of $g$ and $E$.
\end{itemize} 
\subsection{$\Omega\neq 0$}

 To find the modulation instability (MI) threshold, we first define the critical frequency ($\Omega_c$) as the frequency at which $\lambda_{\pm}$ has a maximum. The value of $\Omega_c$ can then be readily found by solving the equation $\frac{d\lambda_\pm(\Omega)}{d\Omega}=0$, and verifies $\Omega^2_c=2I-\Delta$. The MI intensity threshold  verifies the condition $\lambda_{\pm}(\Omega_c)=0$, which leads to:  
\begin{equation}
    I_{th}=\frac{-(1-\xi)+\sqrt{(1-\xi)^2+4\xi(1-\mathcal{G}_0)}}{2\xi}.
    \label{eq:Ith}
\end{equation}
Therefore, the CW solution is unstable against periodic perturbations when the conditions $I>I_{th}$ and $2I_{th}>\Delta$ are satisfied simultaneously and stable otherwise.

Eigenvalues of with two equations (characteristic polynomial)

\begin{align}
    &-\lambda^3+(2(g-1) - \mu ( 1+  \xi I) )\lambda^2+ (-1 - \Delta^2 - 2 \mu + g (2 - g + 2 \mu) + 2 I(\Delta + (-2 + g) \mu \xi])\lambda- \\
    &\mu [(-1 + g)^2 + \Delta^2 -2 (-1 + \Delta)\xi I^2 +I(-2 \Delta + (3 + (-4 + g) g + \Delta^2) \xi) ]=0
\end{align}